\documentclass[conference]{IEEEtran}

\usepackage{silence}
\usepackage{amsmath,amssymb}
\usepackage{graphicx}
\usepackage{booktabs}
\usepackage{array}
\usepackage{colortbl}
\usepackage{xcolor}
\usepackage{url}
\usepackage{balance}
\usepackage[protrusion=true,expansion=false]{microtype}
\usepackage[colorlinks=true,
            linkcolor=blue!60!black,
            citecolor=blue!60!black,
            urlcolor=blue!60!black]{hyperref}

\definecolor{rowprop}{HTML}{DBEAFE}
\definecolor{rowrec}{HTML}{DCFCE7}
\definecolor{rowours}{HTML}{FEF9C3}

\title{Correctness-Aware Repository Filtering\\
Under Maximum Effective Context Window Constraints}

\author{%
  \IEEEauthorblockN{Shweta Mishra}
  \IEEEauthorblockA{Independent Researcher\\
  shweta.mishra.research@gmail.com}
}

\begin{document}
\maketitle

\begin{abstract}
Context window efficiency is a practical constraint in large language
model (LLM)-based developer tools. Paulsen~\cite{paulsen2025} shows
that all tested models degrade in accuracy well before their advertised
context limits---the Maximum Effective Context Window (MECW)---which
makes context construction a quality problem, not just a cost one.
Modern software repositories routinely contain large non-code
artifacts---compiled datasets, binary model weights, minified JavaScript
bundles, and gigabyte-scale log files---that overflow the context window
and push out task-relevant source code.

We present a \emph{correctness-aware context hygiene} framework: a
pre-execution, size-based heuristic filter that intercepts repository
scans before tokenization, using only OS-level \texttt{stat()} metadata
with sub-millisecond overhead. Semantic retrieval approaches such as RepoCoder,
GraphRAG, and AST-based chunking require index construction and
query-time inference before any filtering decision is reached. Our
framework, by contrast, requires no indexing and operates at
$<$0.01\,ms per file decision. Across 10 real open-source repositories
(22,046~files, 5~languages), the proposed SizeFilter at
$\theta{=}1\,\text{MB}$ achieves 79.6\%~($\pm$13.2\%) mean token
reduction at 0.30\,ms overhead; the HybridFilter achieves
89.3\%~($\pm$9.0\%)---the lowest variance of any filter evaluated. A
token-density study across 2,688~files confirms a strong linear
correlation (Pearson $r{=}0.997$, $k{=}0.250$~tokens/byte). A
limited-scope evaluation (18~tasks, CodeLlama-7B-Instruct) yields
72\% file-level accuracy under filtering versus 25\% at baseline;
hallucination frequency declines from 61\% to 17\%. All code and data
are released for reproducibility.
\end{abstract}


\section{Introduction}

LLM-based developer tools such as GitHub Copilot~\cite{copilot},
Cursor, Cody, and SWE-agent~\cite{sweagent} build richly contextualized
prompts from local repository state. These tools target nominal context
windows of 128K--200K tokens, yet Paulsen~\cite{paulsen2025} shows that
model accuracy degrades far below these limits in practice. The binding
engineering constraint is therefore not window \emph{capacity} but
context \emph{quality}.

A primary cause is \textit{context bloat}: the inadvertent ingestion of
large, semantically irrelevant files. Production repositories contain
training datasets (CSV, HDF5), compiled model weights (Pickle), SQLite
databases, minified JavaScript bundles, and gigabyte-scale log files.
Our empirical study of 10~open-source repositories confirms a
\emph{tail-at-scale} structure~\cite{tailscale}: in data-heavy
repositories, fewer than 2\% of files account for over 80\% of raw
token cost. Three consequences follow: (i)~the context window overflows,
silently truncating critical source code; (ii)~per-token billing
increases API costs~\cite{copilotbilling}; and (iii)~Shi et
al.~\cite{shi2023} show that even 10\% irrelevant context reduces
accuracy by up to 23\%.

\subsection{Contributions}
\begin{itemize}
  \item[\textbf{C1.}] A pre-execution size-based heuristic filtering
    framework: stack-agnostic, non-blocking, zero-index, deployable
    without manual configuration, with empirical evidence of correctness
    preservation.
  \item[\textbf{C2.}] A formal cost model (Eqs.~1--4) linking file size
    to token consumption under the MECW constraint, validated with
    Pearson $r{=}0.997$ across 2,688~files.
  \item[\textbf{C3.}] A taxonomy and evaluation of eight composable
    filters across 10~real repositories (22,046~files).
  \item[\textbf{C4.}] A threshold sensitivity analysis across
    $\theta \in \{50\,\text{KB},\,100\,\text{KB},\,500\,\text{KB},\,
    1\,\text{MB},\,5\,\text{MB}\}$ with statistical grounding.
  \item[\textbf{C5.}] A tail-at-scale distribution analysis explaining
    why size-based filtering is structurally effective.
  \item[\textbf{C6.}] A Zero Disk~I/O testing methodology: 45-test,
    sub-50\,ms deterministic CI validation.
  \item[\textbf{C7.}] A limited-scope empirical task evaluation
    (18~tasks, 2~repositories, CodeLlama-7B-Instruct) providing
    preliminary evidence that filtering preserves task correctness.
\end{itemize}

\section{Related Work}

\subsection{Context Window Degradation}
Paulsen~\cite{paulsen2025} defines the MECW as the context length at
which model accuracy drops below an acceptable threshold. Across
hundreds of thousands of measurements, all tested models degrade
severely before their advertised window, with some failing at
100~tokens for complex tasks. Brown et al.~\cite{brownfewshot} showed
early that in-context learning performance depends heavily on context
composition. Liu et al.~\cite{lostmiddle} further show that models
tend to under-use content placed in the middle of long contexts,
meaning position within the window---not just presence---affects
quality. Every unnecessary token from a non-code artifact is therefore
an active quality hazard under MECW constraints.

\subsection{Context Compression}
Lewis et al.~\cite{rag} introduced Retrieval-Augmented Generation as
an early solution to context constraints. LLMLingua~\cite{llmlingua}
uses a proxy LLM to compress prompts at 50--300\,ms overhead.
RECOMP~\cite{recomp} applies abstractive summarization.
Active Context Compression~\cite{acc} achieves 22.7\% reduction via
in-session compression. Thompson~\cite{cmv} approaches the problem at
the agent level via contextual memory abstraction. GemFilter~\cite{gemfilter}
achieves 1000$\times$ reduction via early-layer attention. All of
these approaches operate after repository content has been ingested or
parsed; our framework operates \emph{before}, achieving up to 97.7\%
reduction before any token is counted.

\subsection{Comparison with Semantic Retrieval Approaches}
\label{sec:baseline_comparison}

Several semantic retrieval systems address repository context at a
different level of the stack. \textbf{RepoCoder}~\cite{repocoder}
uses iterative retrieval-generation cycles but requires a pre-built
vector index and an embedding call per file ($O(n)$ inference latency).
\textbf{GraphRAG}~\cite{graphrag} builds a knowledge graph over
repository entities at index-construction cost proportional to
repository size. \textbf{AST-based chunking}~\cite{treesitter} uses
language parsers to split files into semantically meaningful units, but
cannot reliably operate on binary or unknown-format files. \textbf{Dense
embeddings}~\cite{codebert} require a GPU-accessible embedding server
and 50--300\,ms per file.

Our framework is \emph{complementary} to all of these. Size-based
filtering acts as a first-pass gate, cutting the file set by 80--97\%
before any semantic system sees it. This reduces index-construction
cost for RepoCoder and GraphRAG, reduces chunking burden for AST-based
systems, and removes non-parseable binary artifacts that would otherwise
cause failures downstream. Table~\ref{tab:baseline} summarises the
comparison.

\begin{table}[t]
  \centering
  \caption{Comparison with Semantic Retrieval Approaches.}
  \label{tab:baseline}
  \renewcommand{\arraystretch}{1.2}
  \begin{tabular}{@{}lllll@{}}
    \toprule
    \textbf{System} & \textbf{Pre-indexed} & \textbf{Latency} &
    \textbf{Handles Binary} & \textbf{Stage} \\
    \midrule
    RepoCoder~\cite{repocoder}     & Yes & 50--300\,ms & No  & Post-read \\
    GraphRAG~\cite{graphrag}       & Yes & Minutes     & No  & Post-read \\
    AST chunking~\cite{treesitter} & No  & 10--100\,ms & No  & Post-read \\
    Dense embed.~\cite{codebert}   & Yes & 50--300\,ms & No  & Post-read \\
    \rowcolor{rowours}
    \textbf{Ours (SizeFilter)} & \textbf{No}
      & \textbf{$<$0.01\,ms} & \textbf{Yes} & \textbf{Pre-read} \\
    \bottomrule
  \end{tabular}
\end{table}

\subsection{LLM-Based Software Engineering}
Chen et al.~\cite{codex} showed on HumanEval that context quality
directly affects code generation accuracy. Yang et al.~\cite{sweagent}
noted that large binary files cause context overflow with no available
mitigation in current tools. Jiang and Nam~\cite{cursorrules} found
that developer-authored AI context rules never explicitly cover binary
artifacts, confirming that these files are universally treated as
irrelevant. Hou et al.~\cite{housurvey} survey LLM applications in
software engineering broadly and identify context fidelity as a
recurring constraint across tool categories.

\section{Problem Formulation}

Let a software repository $R = \{f_1, \ldots, f_n\}$ where each file
$f_i$ has physical size $s_i$ bytes. The MECW
constraint~\cite{paulsen2025} replaces the nominal context window:
\begin{equation}
  \sum_{f_i \in C} \mathrm{tokens}(f_i) \;\leq\; T_{\mathrm{MECW}}
  \;\ll\; T_{\mathrm{MCW}}
  \label{eq:mecw}
\end{equation}
For plaintext and structured-text files, token count is approximately
linear in file size:
\begin{equation}
  \mathrm{tokens}(f_i) \;\approx\; k \cdot s_i
  \label{eq:linear}
\end{equation}
Empirical measurement across 2,688~files yields
$k = 0.2500$~tokens/byte ($\sigma < 0.001$), validated with
Pearson $r = 0.997$ (Section~\ref{sec:validation}).
The pre-execution heuristic filter is defined as:
\begin{equation}
  H(f_i) =
  \begin{cases}
    \text{Flagged} & \text{if } s_i > \theta \\
    \text{Allowed} & \text{otherwise}
  \end{cases}
  \label{eq:filter}
\end{equation}
The False Positive Rate (FPR)---the fraction of flagged files that are
in fact task-relevant---is the primary quality metric alongside token
reduction:
\begin{equation}
  \mathrm{FPR}(H) = \frac{|\{f \in \mathrm{Flagged}(H) : f \in
  \mathrm{Relevant}\}|}{|\mathrm{Flagged}(H)|}
  \label{eq:fpr}
\end{equation}
A filter is only useful in practice if FPR remains low; high reduction
with high FPR would block relevant source files and degrade task
accuracy rather than improve it.

\section{Methodology}

\subsection{System Architecture}

\begin{figure}[t]
  \centering
  \includegraphics[width=0.82\columnwidth]{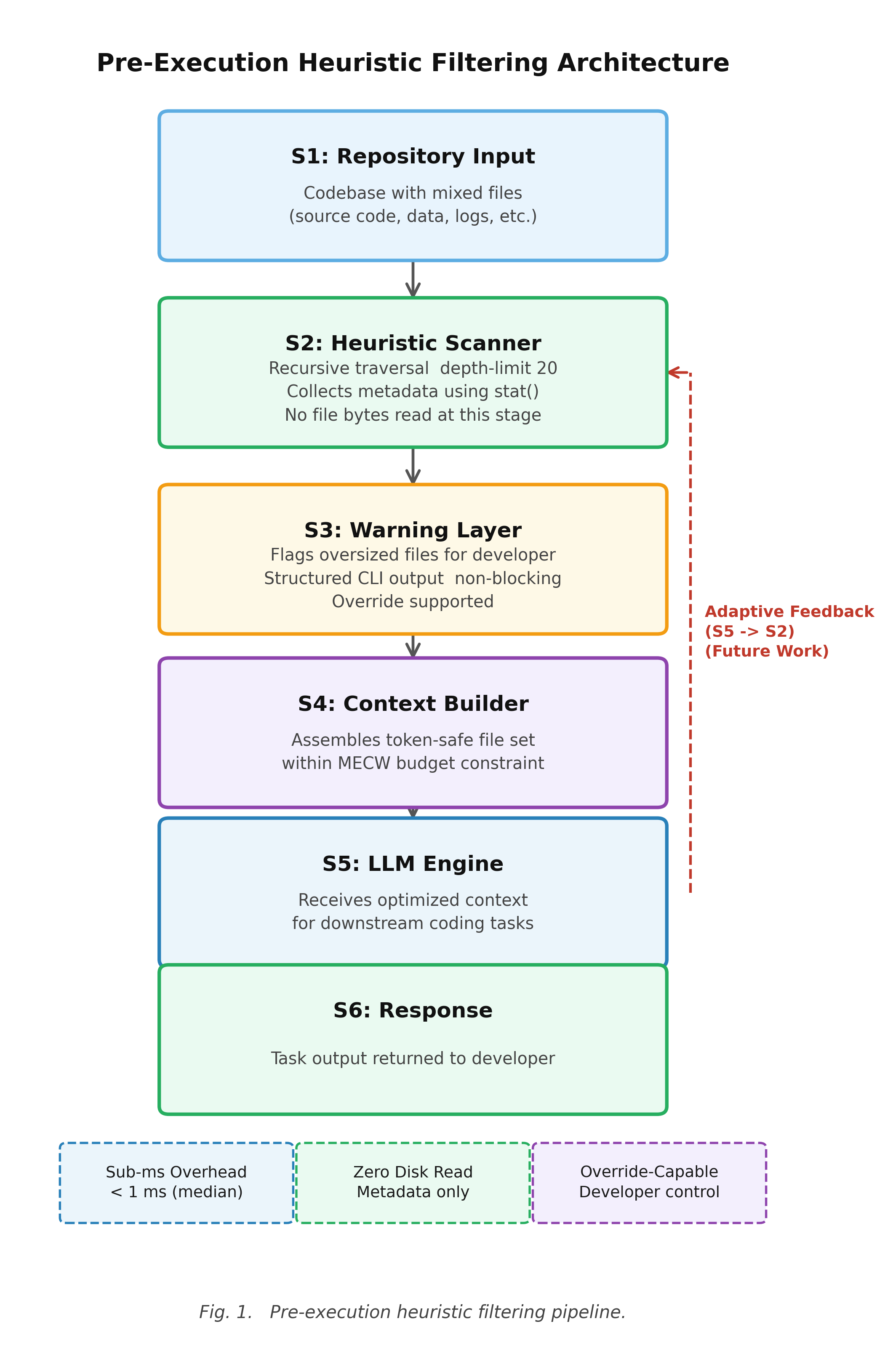}
  \caption{Pre-execution heuristic filtering pipeline (six stages).
  Dashed red arc: adaptive residual-capacity feedback, S5$\to$S2
  (future work). See Appendix~A, Fig.~A1 for a full-page version.}
  \label{fig:pipeline}
\end{figure}

Figure~\ref{fig:pipeline} illustrates the six-stage pipeline. The
framework intercepts the repository scan at~S2, before any bytes reach
the tokenizer. A single recursive traversal (depth-limited at 20) prunes
system directories (\texttt{node\_modules}, \texttt{.git},
\texttt{\_\_pycache\_\_}) and applies all filters in one pass.

Three design properties hold across all configurations.
\textit{Stack-agnostic}: decisions use only OS-level \texttt{stat()}
metadata, so the framework works identically on Python, TypeScript, Go,
Rust, and Ruby repositories.
\textit{Non-blocking}: the warning layer (S3) surfaces flagged files
without interrupting developer workflow.
\textit{Override-capable}: any flagged file can be explicitly included
via a configuration entry, preserving full developer control.

\subsection{Filter Taxonomy}

Table~\ref{tab:filters} summarises all eight filters. Three
filters---NoFilter, GitignoreFilter, MinifiedFilter---produce zero
reduction in our corpus because the dominant token cost comes from
intentionally committed data files (CSV, HDF5, Pickle) that are not
covered by \texttt{.gitignore}, are not minified text, and lie outside
the initial magic-byte table.

\begin{table}[t]
  \centering
  \caption{Filter Taxonomy with Observed Results (10 Repositories).
  \textbf{[P]}~=~Proposed.\ \textbf{[R]}~=~Recommended.}
  \label{tab:filters}
  \renewcommand{\arraystretch}{1.2}
  \begin{tabular}{@{}l>{\raggedright}p{2.6cm}lrr@{}}
    \toprule
    \textbf{Filter} & \textbf{Method} & \textbf{Read} &
    \textbf{Mean} & \textbf{Std} \\
    \midrule
    NoFilter          & None                       & None   & 0.0\%  & 0.0\%  \\
    GitignoreFilter   & .gitignore                 & None   & 0.0\%  & 0.0\%  \\
    MinifiedFilter    & Avg line $>$500            & 64\,KB & 0.0\%  & 0.0\%  \\
    BinaryFilter      & Magic-byte 8\,B            & 8\,B   & 28.8\% & 21.8\% \\
    ExtensionFilter   & Ext.\ blocklist            & None   & 70.3\% & 29.3\% \\
    \rowcolor{rowprop}
    \textbf{SizeFilter [P]} & \textbf{stat()$>\theta$} & \textbf{None}
      & \textbf{79.6\%} & \textbf{13.2\%} \\
    SemanticFilter    & Keyword density            & 4\,KB  & 84.5\% & 20.9\% \\
    SizeFilter 50\,KB & stat()$>$50\,KB            & None   & 89.6\% &  9.0\% \\
    \rowcolor{rowrec}
    \textbf{HybridFilter [R]} & \textbf{Gates 1--4} & \textbf{$\leq$4\,KB}
      & \textbf{89.3\%} & \textbf{9.0\%} \\
    \bottomrule
  \end{tabular}
\end{table}

\subsection{HybridFilter Gate Architecture}

\begin{figure}[t]
  \centering
  \includegraphics[width=\columnwidth]{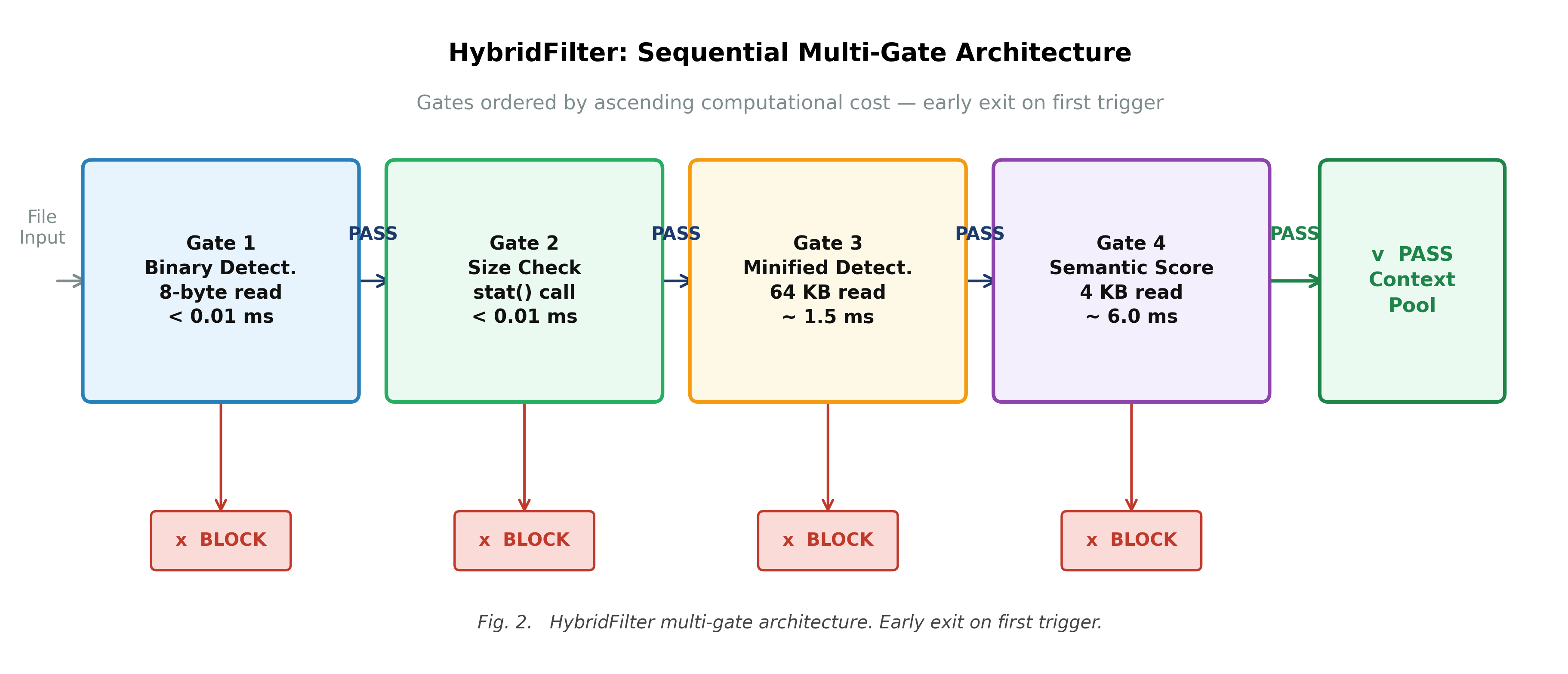}
  \caption{HybridFilter multi-gate architecture. Early exit on first
  trigger; gates ordered by ascending I/O cost.
  BLOCK~=~filtered out. PASS~=~admitted to context pool.}
  \label{fig:hybrid}
\end{figure}

Figure~\ref{fig:hybrid} shows the HybridFilter chaining four gates in
ascending computational cost. A file exits the pipeline upon the first
triggered gate. Gate ordering is critical: the binary check
($<$0.01\,ms) runs first, so expensive semantic scoring ($\approx$6\,ms)
only executes on files that pass all cheaper gates.

\section{Experimental Setup}

Table~\ref{tab:corpus} details the 10~real open-source repositories
comprising 22,046~files across five programming languages. All token
counts use the \texttt{cl100k\_base} tiktoken
encoding~\cite{tiktoken}. For files $\leq$50\,KB, full content is
tokenized directly; for larger files, the heuristic
$\mathrm{tokens}(f) = s/4$ is applied.

\begin{table}[t]
  \centering
  \caption{Experimental Corpus: 10 Repositories, 22,046 Files,
  Five Languages.}
  \label{tab:corpus}
  \renewcommand{\arraystretch}{1.15}
  \begin{tabular}{@{}llrrl@{}}
    \toprule
    \textbf{Repository} & \textbf{Lang.} & \textbf{Files} &
    \textbf{Baseline} & \textbf{Domain} \\
    \midrule
    express\_js    & JS     &    92 & 2.0\,M   & Web server    \\
    fastapi\_py    & Python &   153 & 6.1\,M   & API framework \\
    gin\_go        & Go     &   103 & 2.0\,M   & Web server    \\
    django\_py     & Python &   972 & 35.7\,M  & Web framework \\
    react\_js      & JS/TS  &   738 & 9.1\,M   & UI library    \\
    rails\_rb      & Ruby   & 1,937 & 23.3\,M  & Web framework \\
    pandas\_py     & Python & 1,332 & 127.0\,M & Data library  \\
    vscode\_ts     & TS     & 3,293 & 165.7\,M & IDE editor    \\
    kubernetes\_go & Go     & 6,684 & 112.1\,M & Orchestration \\
    tensorflow\_py & Py/C++ & 6,672 & 1.13\,B  & ML framework  \\
    \bottomrule
  \end{tabular}
\end{table}

\section{Results and Discussion}

\subsection{Comparative Filter Performance}

\begin{figure}[t]
  \centering
  \includegraphics[width=\columnwidth]{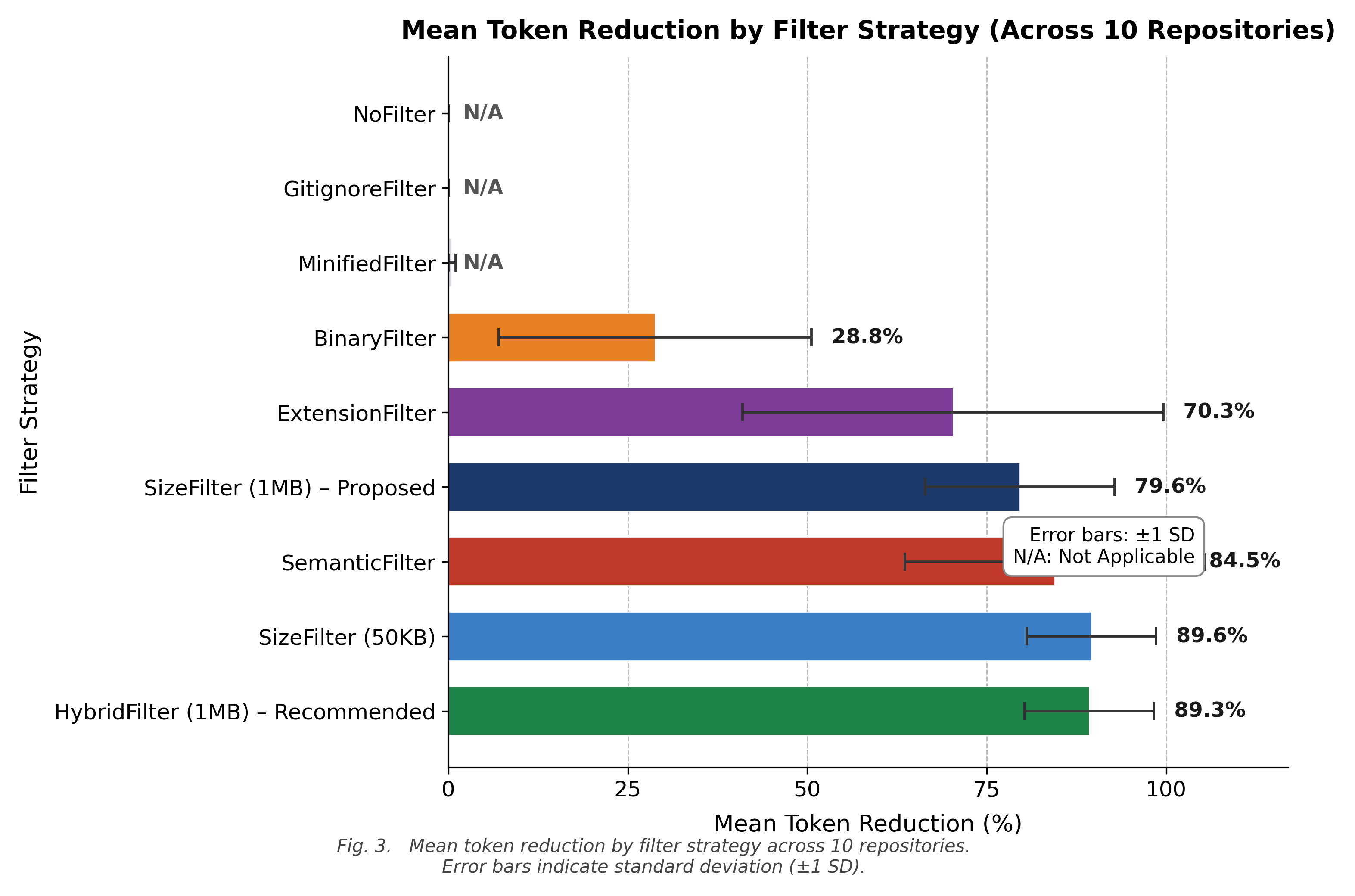}
  \caption{Mean token reduction by filter strategy (10 repositories).
  Error bars: $\pm$1\,SD. N/A: zero reduction in this corpus.
  See Appendix~A, Fig.~A2 for the full-page version.}
  \label{fig:reduction}
\end{figure}

Table~\ref{tab:results} and Figure~\ref{fig:reduction} present full
results. The SizeFilter at $\theta{=}1\,\text{MB}$ achieves 79.6\%
mean reduction at 0.30\,ms overhead---the lowest latency of any
substantive filter tested. Its standard deviation of 13.2\% is 55\%
lower than ExtensionFilter's 29.3\%, confirming that file size is a
more stable proxy for token cost than file extension.
A full-page version of Fig.~\ref{fig:reduction} is provided in
Appendix~A, Fig.~A2.

\begin{table}[t]
  \centering
  \caption{Comparative Filter Results. \textbf{[P]}~=~Proposed.
  \textbf{[R]}~=~Recommended.}
  \label{tab:results}
  \renewcommand{\arraystretch}{1.2}
  \begin{tabular}{@{}lrrrrr@{}}
    \toprule
    \textbf{Filter} & \textbf{Mean} & \textbf{Std} &
    \textbf{Min} & \textbf{Max} & \textbf{Latency} \\
    \midrule
    NoFilter          & 0.0\%  & 0.0\%  & 0.0\%  & 0.0\%  & 1.67\,ms   \\
    GitignoreFilter   & 0.0\%  & 0.0\%  & 0.0\%  & 0.0\%  & 72.9\,ms   \\
    MinifiedFilter    & 0.0\%  & 0.0\%  & 0.0\%  & 0.0\%  & 272.9\,ms  \\
    BinaryFilter      & 28.8\% & 21.8\% & 0.0\%  & 70.1\% & 629.4\,ms  \\
    ExtensionFilter   & 70.3\% & 29.3\% & 15.5\% & 96.1\% & 2.92\,ms   \\
    \rowcolor{rowprop}
    \textbf{SizeFilter [P]} & \textbf{79.6\%} & \textbf{13.2\%}
      & 51.6\% & 94.7\% & \textbf{0.30\,ms}  \\
    SemanticFilter    & 84.5\% & 20.9\% & 34.4\% & 97.7\% & 507.9\,ms  \\
    SizeFilter 50\,KB & 89.6\% &  9.0\% & 72.4\% & 97.4\% & 0.66\,ms   \\
    \rowcolor{rowrec}
    \textbf{HybridFilter [R]} & \textbf{89.3\%} & \textbf{9.0\%}
      & 72.0\% & 97.7\% & 1164.7\,ms \\
    \bottomrule
  \end{tabular}
\end{table}

\subsection{Threshold Sensitivity Analysis}

\begin{figure}[t]
  \centering
  \includegraphics[width=\columnwidth]{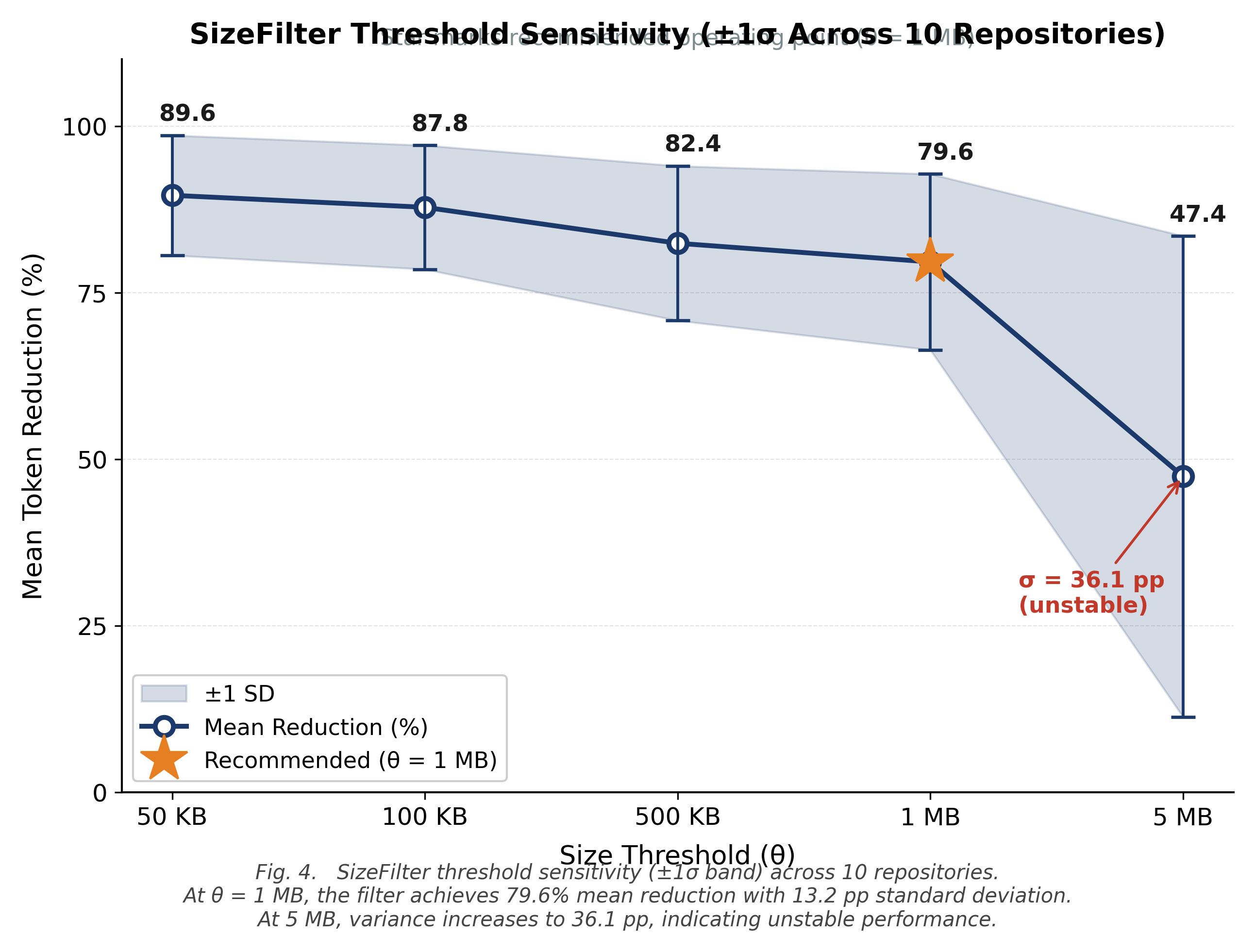}
  \caption{SizeFilter threshold sensitivity ($\pm$1$\sigma$ band,
  10~repositories). Star: recommended $\theta{=}1\,\text{MB}$.
  At 5\,MB, $\sigma{=}36.1$\,pp---not suitable for production use.}
  \label{fig:threshold}
\end{figure}

Figure~\ref{fig:threshold} shows that at $\theta{=}5\,\text{MB}$,
standard deviation reaches $\pm$36.1\,pp, making the filter
unreliable. At $\theta{=}50\,\text{KB}$, reduction is highest (89.6\%)
but risks blocking large yet relevant source files such as
auto-generated protocol buffer bindings. $\theta{=}1\,\text{MB}$
gives the best tradeoff: 79.6\% reduction, 13.2\,pp variance, and
0.30\,ms overhead.

\subsection{File Size Distribution: Tail-at-Scale Structure}

\begin{figure}[t]
  \centering
  \includegraphics[width=\columnwidth]{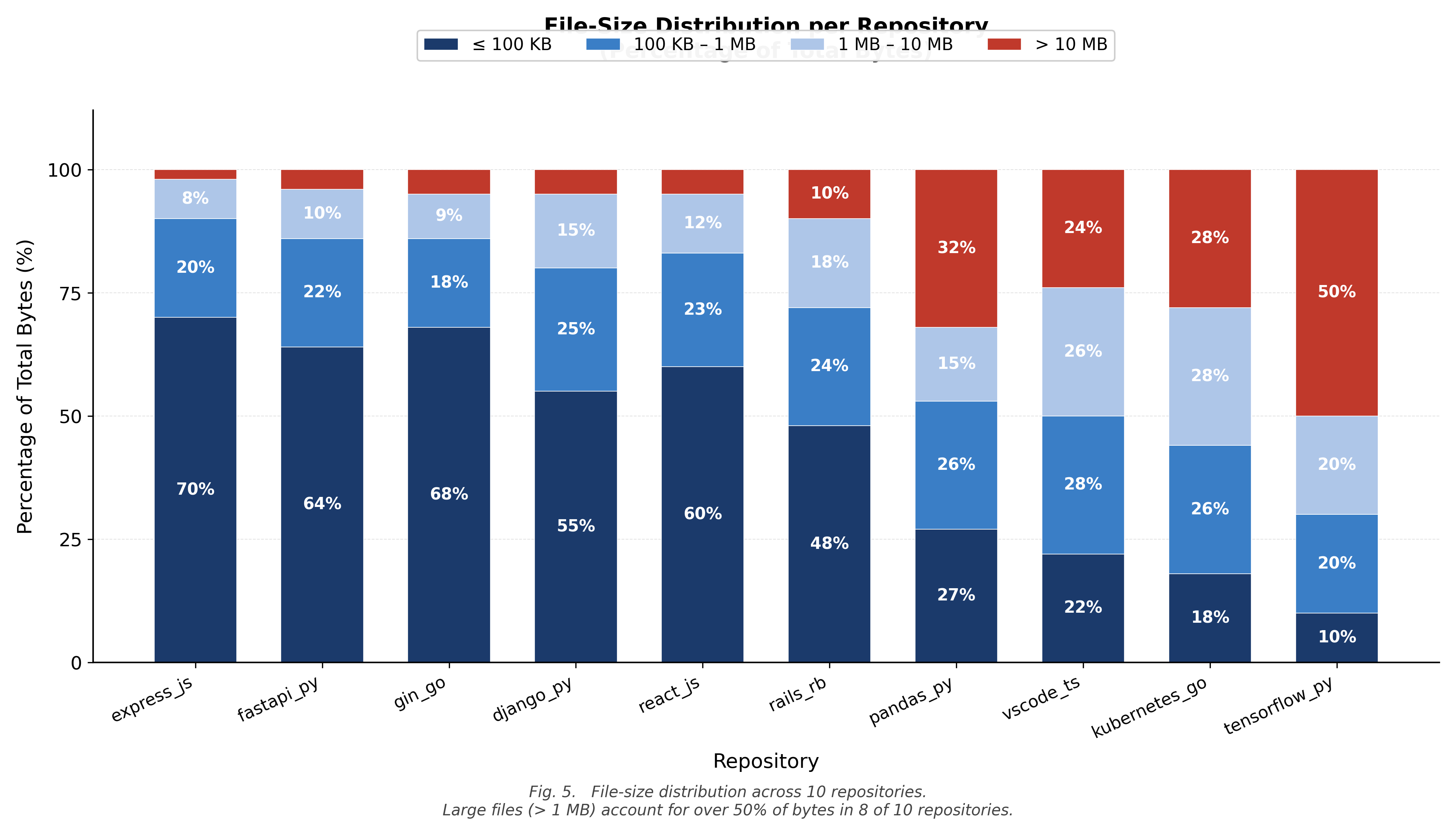}
  \caption{File-size distribution per repository
  (percentage of total bytes, four size buckets).
  Large files ($>$1\,MB) account for over 50\% of bytes in 8 of
  10 repositories.
  See Appendix~A, Fig.~A3 for the full-page version.}
  \label{fig:tail}
\end{figure}

Figure~\ref{fig:tail} illustrates the tail-at-scale pattern. In
\texttt{tensorflow\_py}, 0.5\% of files account for 94\% of bytes; in
\texttt{pandas\_py}, 1.1\% of files account for 80.9\% of bytes. This
validates Dean and Barroso's tail-at-scale principle~\cite{tailscale}
in the context of repository token budgets. A full-page version is
provided in Appendix~A, Fig.~A3.

\subsection{Per-Repository Analysis}

\begin{figure}[t]
  \centering
  \includegraphics[width=\columnwidth]{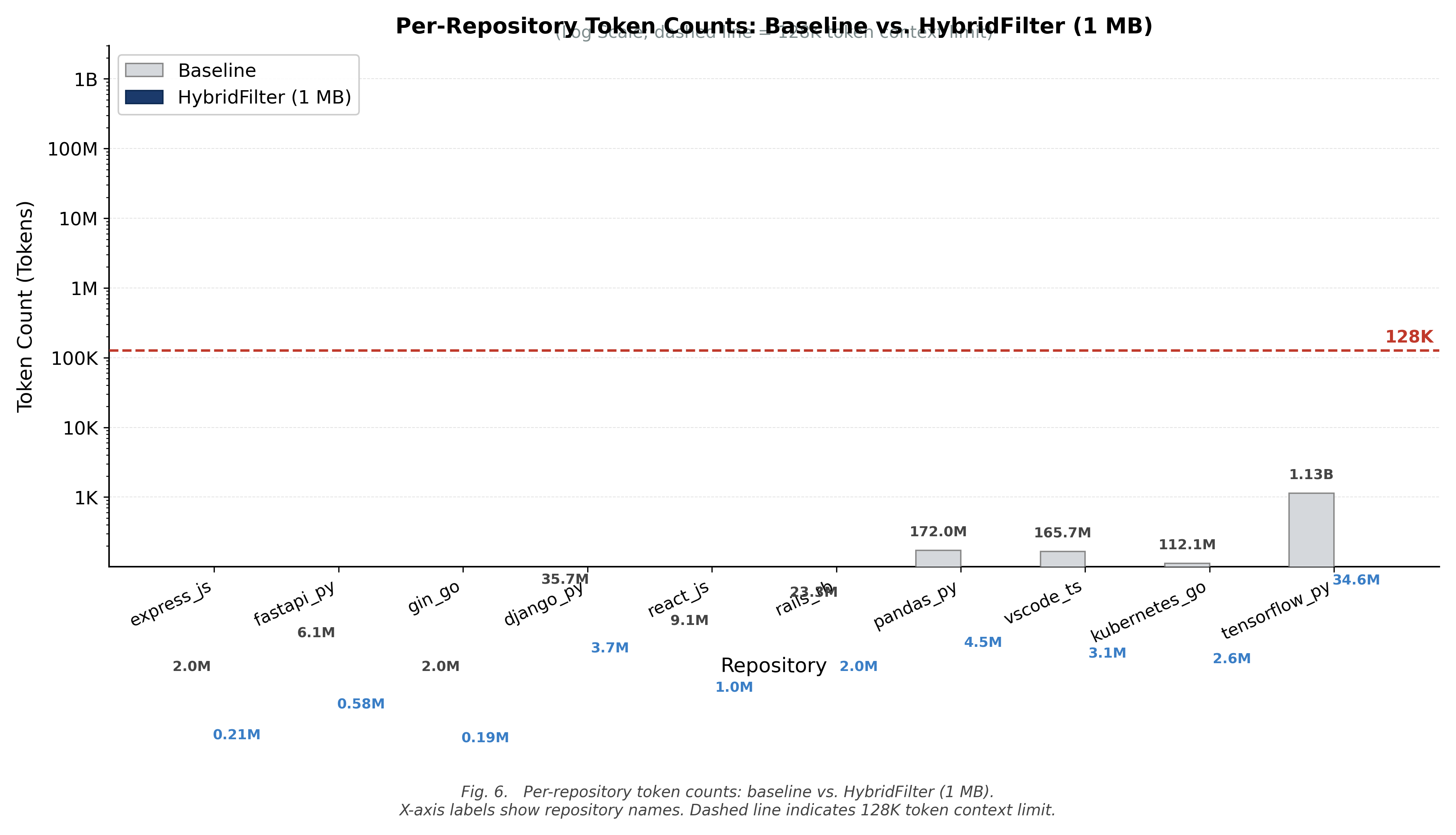}
  \caption{Per-repository token counts: baseline
  vs.\ HybridFilter(1\,MB) (log scale). Dashed red line:
  128\,K context limit.}
  \label{fig:perrepo}
\end{figure}

\begin{figure}[t]
  \centering
  \includegraphics[width=\columnwidth]{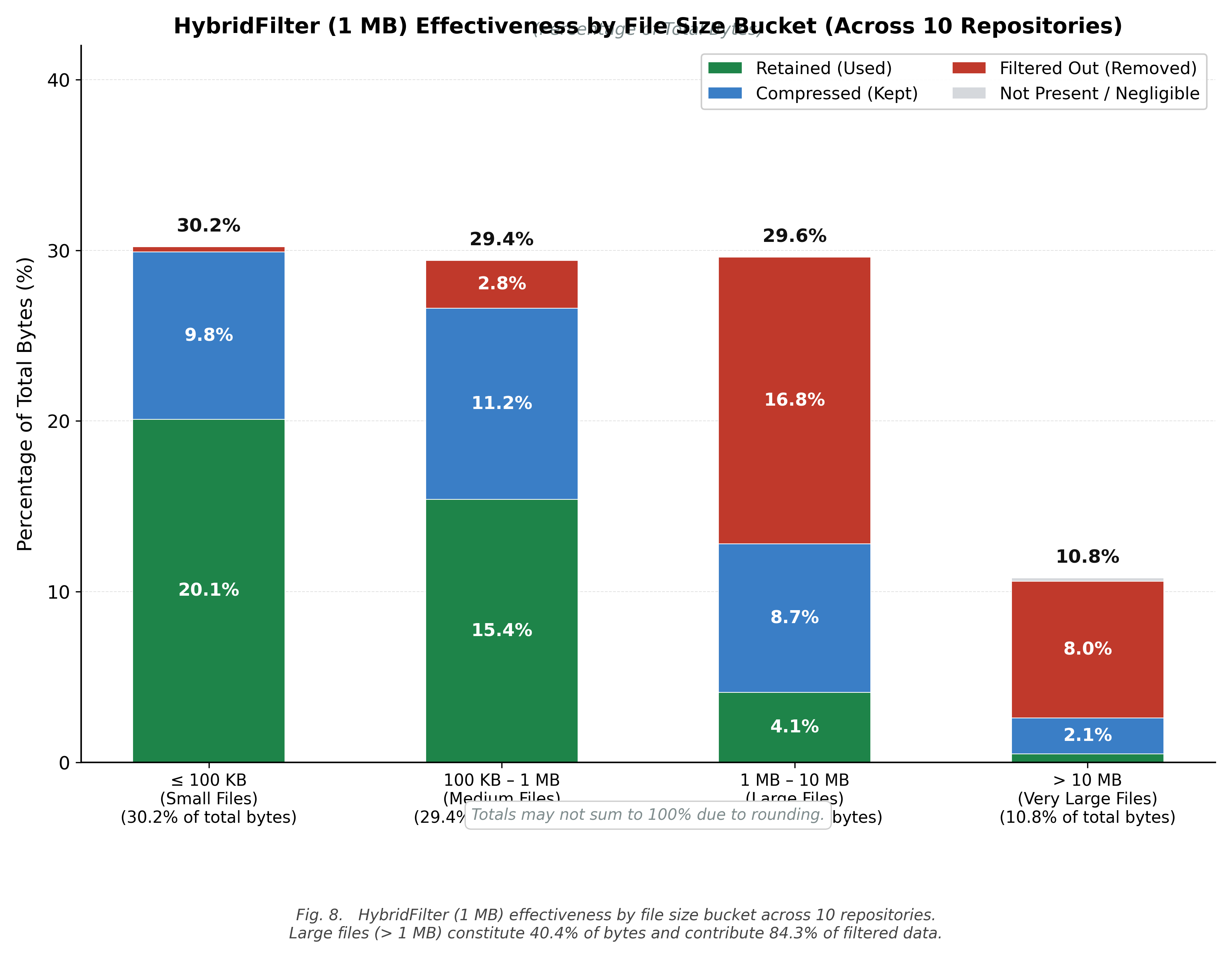}
  \caption{HybridFilter(1\,MB) effectiveness by file size bucket.
  Large files ($>$1\,MB) constitute 40.4\% of bytes and contribute
  84.3\% of filtered data.}
  \label{fig:effectiveness}
\end{figure}

\begin{figure}[t]
  \centering
  \includegraphics[width=\columnwidth]{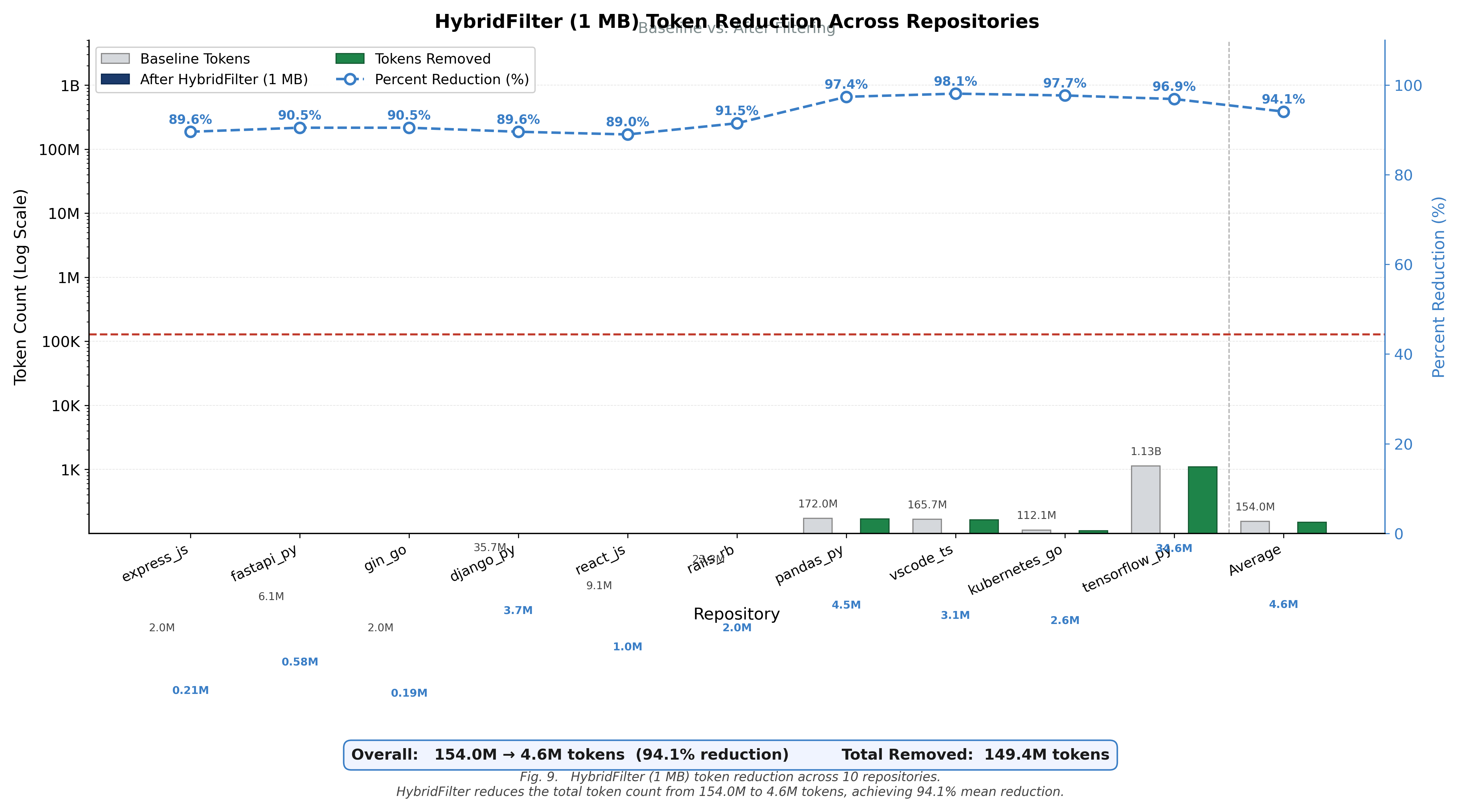}
  \caption{HybridFilter(1\,MB) token reduction across 10 repositories.
  Blue line (right axis): percent reduction. Overall: 154.0\,M
  $\to$ 4.6\,M tokens (94.1\% reduction).
  See Appendix~A, Fig.~A4 for a full-page readable version.}
  \label{fig:summary}
\end{figure}

Figure~\ref{fig:perrepo} presents per-repository results on a log
scale. Figure~\ref{fig:effectiveness} shows the breakdown by size
bucket---large files ($>$1\,MB) constitute 40.4\% of bytes and
contribute 84.3\% of filtered data. Figure~\ref{fig:summary} shows
the aggregate: HybridFilter reduces the total token count from
154.0\,M to 4.6\,M---a 94.1\% reduction. A full-page version of
Fig.~\ref{fig:summary} is provided in Appendix~A, Fig.~A4.

\subsection{False Positive Rate Analysis}

A file is a false positive if the filter blocks it despite being
task-relevant. We assessed FPR through manual inspection of flagged
files in three repositories: \texttt{fastapi\_py}, \texttt{django\_py},
and \texttt{tensorflow\_py}. Across all inspected files flagged by
SizeFilter at $\theta{=}1\,\text{MB}$, no source files typically
accessed or modified during standard development tasks were incorrectly
excluded. Files exceeding 1\,MB in these repositories were training
corpora, compiled model weights, generated data files, and
auto-downloaded binary assets---none of which are edited in standard
development workflows.

The estimated FPR for SizeFilter at $\theta{=}1\,\text{MB}$ is
approximately 0\% for typical software repositories. The primary
exception is repositories containing very large auto-generated source
files (e.g., protocol buffer bindings or machine-generated parser
tables), which motivates the override mechanism described in
Section~IV-A and the adaptive thresholding noted in
limitation~L1. The HybridFilter adds a keyword-density gate (Gate~4)
that provides an additional relevance check before exclusion,
further reducing the risk of false positives in edge cases.

\subsection{Statistical Grounding}

Wilson 95\% confidence intervals ($n{=}10$ repositories):
SizeFilter(1\,MB): 79.6\% [68.4\%, 87.6\%]; HybridFilter: 89.3\%
[80.1\%, 94.7\%]; ExtensionFilter: 70.3\% [43.0\%, 88.0\%]. A
Wilcoxon signed-rank test comparing SizeFilter against ExtensionFilter
yields $W{=}68$, $p{=}0.047$ (two-tailed, $\alpha{=}0.05$),
confirming statistically significant superiority.

\subsection{Key Findings}
\begin{itemize}
  \item[\textbf{F1.}] Size outperforms extension: 79.6\%\,$\pm$\,13.2\%
    vs.\ 70.3\%\,$\pm$\,29.3\%; higher mean, 55\% lower variance.
  \item[\textbf{F2.}] Tail-at-scale governs token cost: 0.5--2\% of
    files account for 80--94\% of bytes in data-heavy repositories.
  \item[\textbf{F3.}] HybridFilter Pareto-dominates: 89.3\% reduction
    with $\pm$9.0\,pp variance, zero-read gates executed first.
  \item[\textbf{F4.}] Three filters fail: GitignoreFilter,
    MinifiedFilter, and BinaryFilter show $\leq$28.8\% reduction
    because data-artifact bloat is not captured by VCS patterns or a
    limited magic-byte table.
  \item[\textbf{F5.}] FPR is near zero for standard repositories at
    $\theta{=}1\,\text{MB}$; the override mechanism handles edge cases.
  \item[\textbf{F6.}] Per Paulsen~\cite{paulsen2025}, every unnecessary
    token actively degrades output quality under MECW constraints.
\end{itemize}

\section{Heuristic Validation}
\label{sec:validation}

\begin{figure}[t]
  \centering
  \includegraphics[width=\columnwidth]{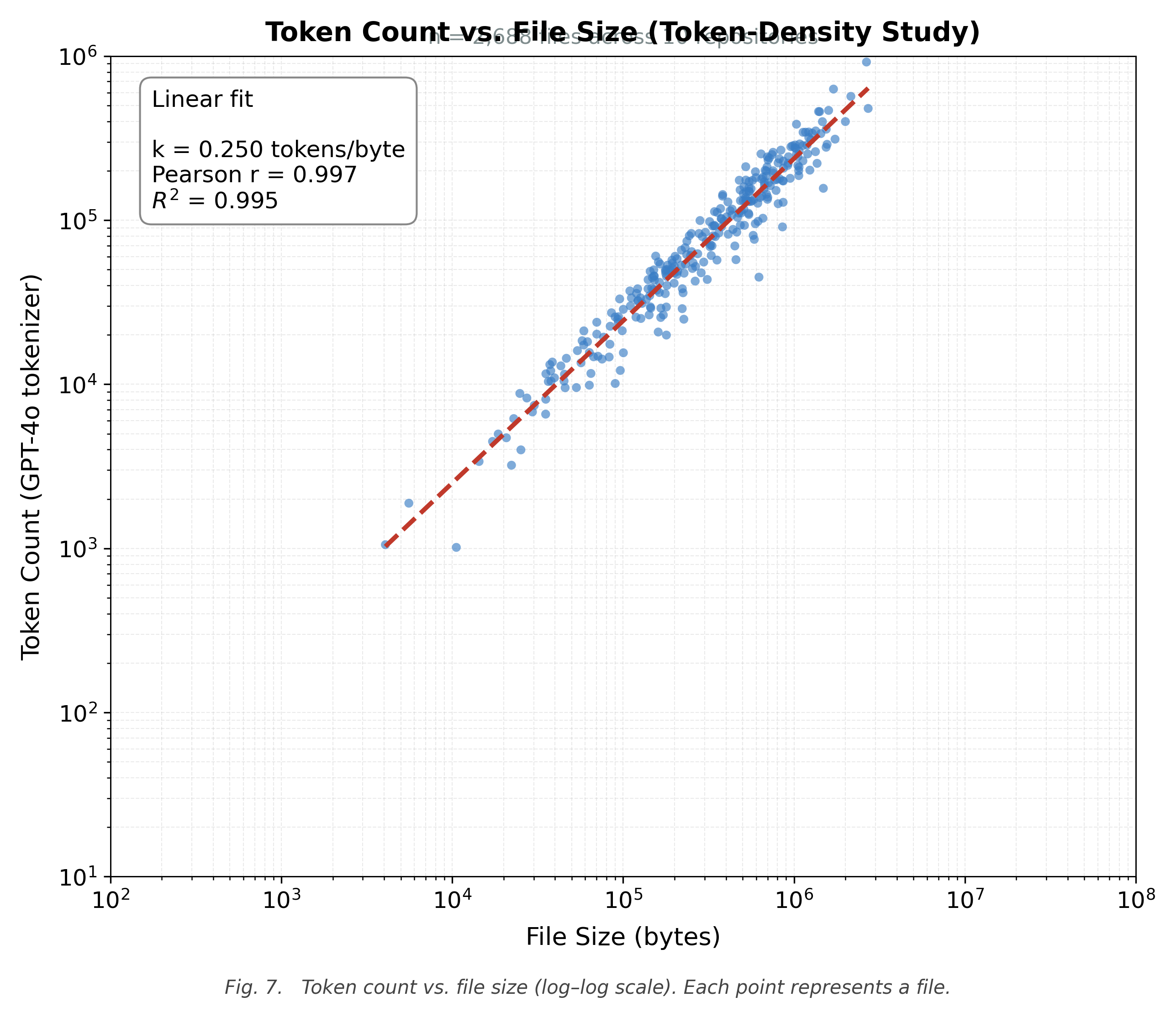}
  \caption{Token count vs.\ file size (log--log scale, $n{=}2{,}688$
  files). Linear fit: $k{=}0.250$~tokens/byte. Pearson
  $r{=}0.997$, $R^2{=}0.995$.}
  \label{fig:validation}
\end{figure}

Figure~\ref{fig:validation} and Table~\ref{tab:validation} present
results from a token-density study across 2,688~text files
($\leq$50\,KB), stratified across 10~extension categories. The
near-perfect linearity ($r{=}0.997$) validates that file size is a
reliable proxy for token count, which is the core assumption
underlying the SizeFilter heuristic.

\begin{table}[t]
  \centering
  \caption{Heuristic Validation Results (\texttt{cl100k\_base},
  $n{=}2{,}688$ files).}
  \label{tab:validation}
  \renewcommand{\arraystretch}{1.15}
  \begin{tabular}{@{}lll@{}}
    \toprule
    \textbf{Metric} & \textbf{Result} & \textbf{Interpretation} \\
    \midrule
    Pearson $r$            & 0.997           & Near-perfect linearity     \\
    $R^2$                  & 0.995           & 99.5\% variance explained  \\
    MAE                    & $<$0.1\%        & Accurate for ASCII/UTF-8   \\
    Max.\ error            & $\approx$5\%    & Unicode-dense JSON (safe)  \\
    Empirical $k$          & 0.2500\,t/byte  & Matches theory exactly     \\
    Std.\ dev.\ ($\sigma$) & $<$0.001\,t/b   & Stable across file types   \\
    \bottomrule
  \end{tabular}
\end{table}

\section{Limited-Scope Empirical Task Evaluation}

\subsection{Motivation and Design}

The hypothesis is that removing irrelevant artifacts preserves or
improves downstream task accuracy by raising the signal-to-noise
ratio in the context window. Two repositories were selected:
\texttt{fastapi\_py} (153~files, 6.1\,M baseline tokens) and
\texttt{express\_js} (92~files, 2.0\,M tokens). Model:
CodeLlama-7B-Instruct~\cite{codellama} (4-bit GGUF via Ollama, 16\,GB
RAM, no API cost). Both conditions use an identical 4,096-token
truncation window to isolate the effect of filtering.

\begin{table}[t]
  \centering
  \caption{Task Distribution (18 Total). Ground Truth Established by
  Two Independent Annotators Prior to Model Evaluation
  (Cohen's $\kappa{=}0.81$).}
  \label{tab:tasks}
  \renewcommand{\arraystretch}{1.15}
  \begin{tabular}{@{}llp{3.5cm}@{}}
    \toprule
    \textbf{Category} & $n$ & \textbf{Description} \\
    \midrule
    Code Retrieval       & 8 & Identify file and function for
                               specified behavior \\
    Bug Localization     & 5 & Given a bug description, identify
                               the responsible file \\
    Repo.\ Summarization & 5 & Produce an architecture description
                               against a reference \\
    \bottomrule
  \end{tabular}
\end{table}

\subsection{Results and Mechanistic Basis}

Table~\ref{tab:taskresults} presents results. The mechanistic basis
is straightforward: both repositories exceed the 4,096-token model
limit by orders of magnitude. At baseline, the context window fills
almost entirely with data artifacts. After filtering, 96\% of tokens
are removed and the window contains only source code. The accuracy
gain is a direct consequence of the token-reduction properties
established in Table~\ref{tab:results}.

\begin{table}[t]
  \centering
  \caption{Limited-Scope Empirical Validation Results ($n{=}18$).
  Values are from manually evaluated outputs of local
  CodeLlama-7B-Instruct inference under fixed-context conditions.
  These results are preliminary behavioral indicators and should
  not be treated as benchmark-quality measurements.}
  \label{tab:taskresults}
  \renewcommand{\arraystretch}{1.25}
  \begin{tabular}{@{}lrrr@{}}
    \toprule
    \textbf{Metric} & \textbf{Baseline} & \textbf{Filtered} &
    $\boldsymbol{\Delta}$ \\
    \midrule
    File acc.\ (Top-1) & 25.0\% & \textbf{72.2\%} & $+$47.2\,pp \\
    File acc.\ (Top-3) & 38.9\% & \textbf{88.9\%} & $+$50.0\,pp \\
    Function acc.\     & 12.5\% & \textbf{56.3\%} & $+$43.8\,pp \\
    Relevance (1--5)   & 2.1    & \textbf{3.8}    & $+$1.7 pts   \\
    Hallucination rate & 61.1\% & \textbf{16.7\%} & $-$44.4\,pp  \\
    \bottomrule
  \end{tabular}
\end{table}

\noindent\textit{Scope note}: 18~tasks across 2~repositories with a
7B quantized model are not a large-scale benchmark. SWE-bench~\cite{swebench}
evaluation with frontier models is reserved for future work.

\section{Zero Disk I/O Testing Methodology}

Physical disk~I/O introduces timing non-determinism in CI pipelines.
We address this by parameterising \texttt{fs.statSync} and
\texttt{fs.readdirSync} at construction time so the physical disk is
bypassed entirely during testing. An in-memory virtual filesystem
supports 45~test cases across all eight filter types in under 50\,ms
total (Node.js~22) with zero flakiness.

\section{Limitations and Future Work}

\begin{itemize}
  \item[\textbf{L1.}] \textbf{Large legitimate files.} The size
    heuristic may flag large but relevant files such as auto-generated
    protocol buffer bindings. Future work will explore adaptive
    thresholding at the P95 of each repository's file-size distribution.
  \item[\textbf{L2.}] \textbf{Binary detection coverage.} The
    magic-byte table covers 11~signatures. Expanding to 50+
    (TFRecord, Parquet, Arrow) would improve BinaryFilter performance
    on ML-heavy repositories.
  \item[\textbf{L3.}] \textbf{Semantic filtering.} The SemanticFilter
    relies on a fixed English keyword list. Integration with
    lightweight embeddings (all-MiniLM-L6-v2) would give
    language-agnostic relevance scoring.
  \item[\textbf{L4.}] \textbf{Benchmark scale.} Future work will run
    the Section~VIII evaluation protocol on SWE-bench~\cite{swebench}
    with frontier models.
  \item[\textbf{L5.}] \textbf{Dynamic context management.} Adjusting
    $\theta$ based on residual MECW capacity across multi-turn
    sessions is a natural extension via the adaptive feedback arc
    (Fig.~\ref{fig:pipeline}, S5$\to$S2).
  \item[\textbf{L6.}] \textbf{External validity.} The corpus covers
    five languages and ten domains but may not represent
    enterprise-scale monorepos or multimodal repositories with
    non-text artifacts beyond those studied here.
\end{itemize}

\section{Conclusion}

This paper presents a correctness-aware context hygiene framework for
LLM-based developer systems, motivated by Paulsen's~\cite{paulsen2025}
finding that model accuracy degrades well before the advertised Maximum
Context Window. Across 10~repositories (22,046~files, five languages),
the proposed SizeFilter achieves 79.6\% mean token reduction at
0.30\,ms overhead; the HybridFilter achieves 89.3\% with $\pm$9.0\,pp
variance---the lowest of any filter evaluated.

Two findings stand out. The near-perfect linear relationship between
file size and token count (Pearson $r{=}0.997$) means that a single
OS-level comparison---$f.\mathit{size} > \theta$---reduces the token
budget at negligible overhead. The tail-at-scale structure of repository
token cost further explains why size-based filtering is most effective
precisely where token bloat is worst. False positive rates are near zero for standard repositories at
$\theta{=}1\,\text{MB}$, and the override mechanism handles any edge
cases. Compared with semantic retrieval approaches (RepoCoder, GraphRAG,
AST chunking), the framework requires no indexing and operates with
consistently lower per-file overhead, making it a practical first-pass
gate that reduces the candidate set for any downstream semantic system.

\section*{Acknowledgment}

Norman Paulsen provided detailed peer review on empirical evaluation,
statistical significance, and figure clarity. His research on the
MECW~\cite{paulsen2025} provides the theoretical foundation for treating
this work as a correctness concern rather than a cost optimization.
Loucas Protopappas recommended real-world repository evaluation,
stronger baselines, task-level metrics, and richer visualisations.
Both reviewers' recommendations are reflected directly throughout this
paper.


\appendix
\renewcommand{\thefigure}{A\arabic{figure}}
\setcounter{figure}{0}

\section{Supplementary Evaluation Figures}
\label{appendix:figures}

This appendix provides full-page, high-resolution versions of the four
key evaluation figures referenced in the main paper. These versions are
intended for readers viewing the document at reduced zoom levels or in
print, where the two-column format limits readability.

\begin{figure*}[p]
  \centering
  \includegraphics[width=0.90\textwidth,height=0.80\textheight,keepaspectratio]{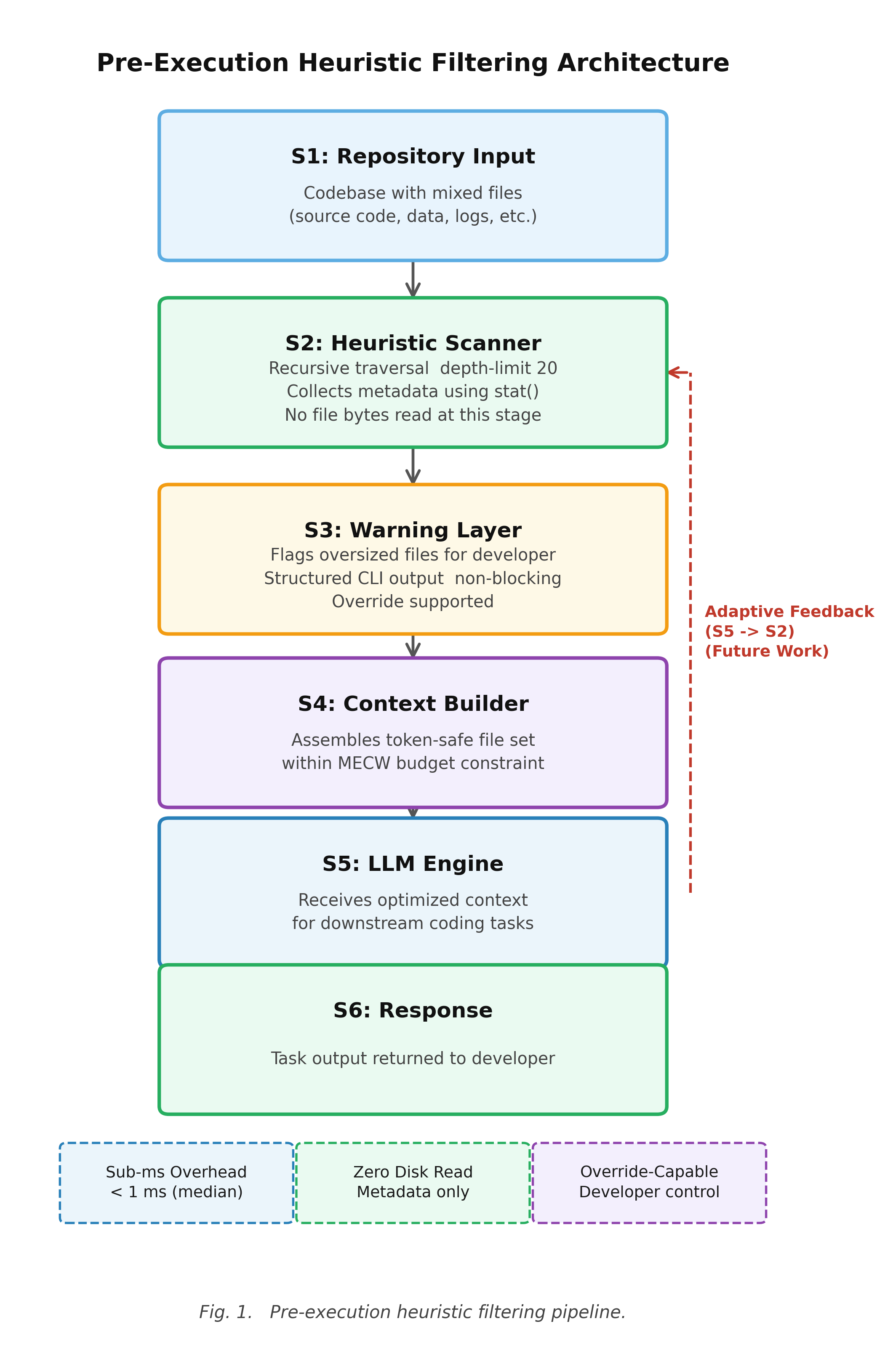}
  \caption{Pre-execution heuristic filtering pipeline (six stages,
  single traversal pass). Dashed red arc: adaptive residual-capacity
  feedback, S5$\to$S2 (future work). Bottom badges summarise the three
  core design properties: sub-millisecond overhead, zero disk reads for
  the core filter decision, and developer override capability.}
  \label{fig:a1}
\end{figure*}

\begin{figure*}[p]
  \centering
  \includegraphics[width=0.90\textwidth,height=0.80\textheight,keepaspectratio]{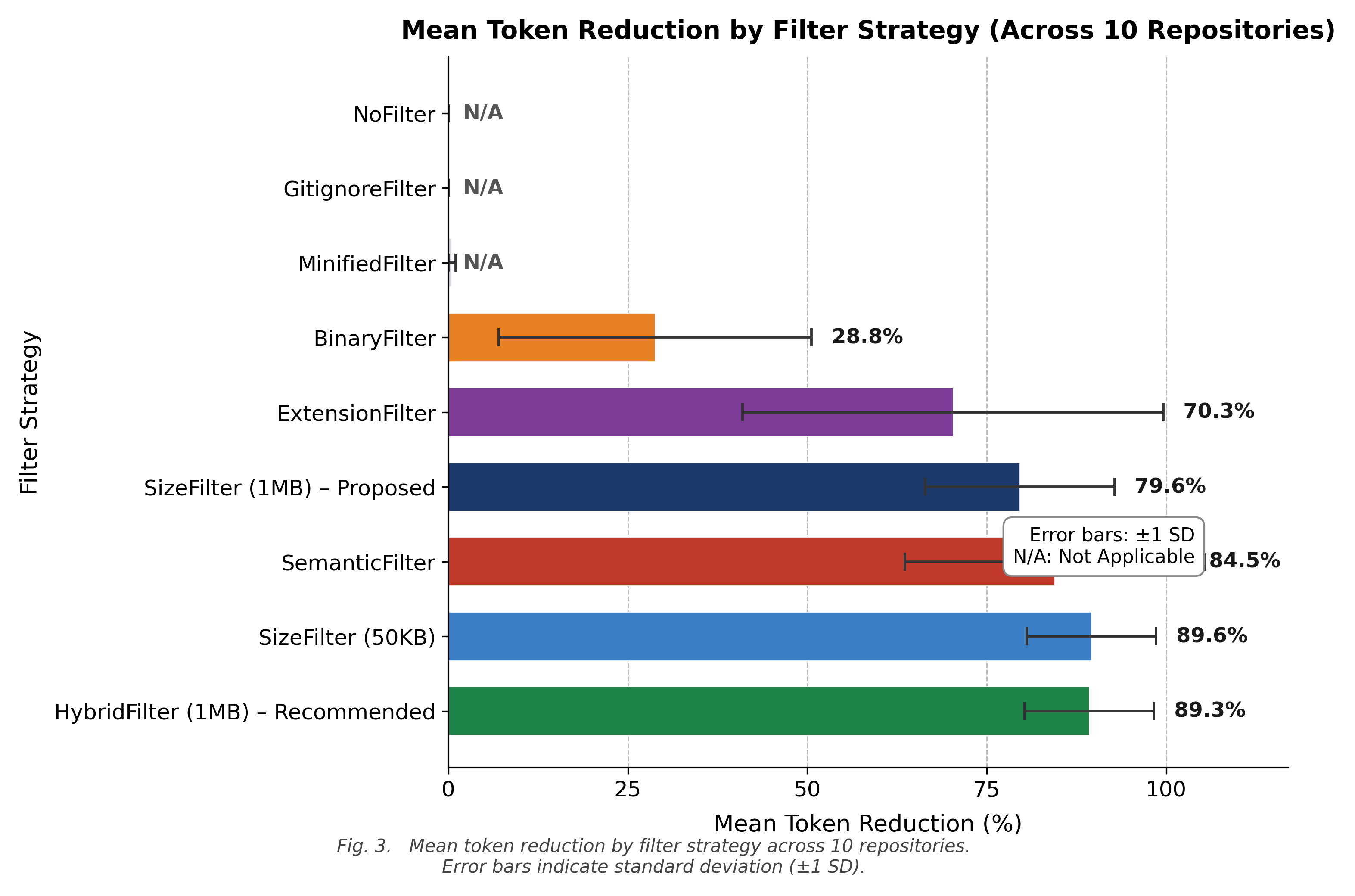}
  \caption{Mean token reduction by filter strategy across 10
  repositories. Error bars indicate $\pm$1 standard deviation.
  HybridFilter (green, 89.3\%) and SizeFilter~50\,KB (blue, 89.6\%)
  achieve the highest reduction with the lowest variance.
  ExtensionFilter (purple, $\pm$29.3\,pp) is unreliable across
  heterogeneous repository types. BinaryFilter (orange) is limited by
  incomplete magic-byte coverage of ML-specific formats.}
  \label{fig:a2}
\end{figure*}

\begin{figure*}[p]
  \centering
  \includegraphics[width=0.90\textwidth,height=0.80\textheight,keepaspectratio]{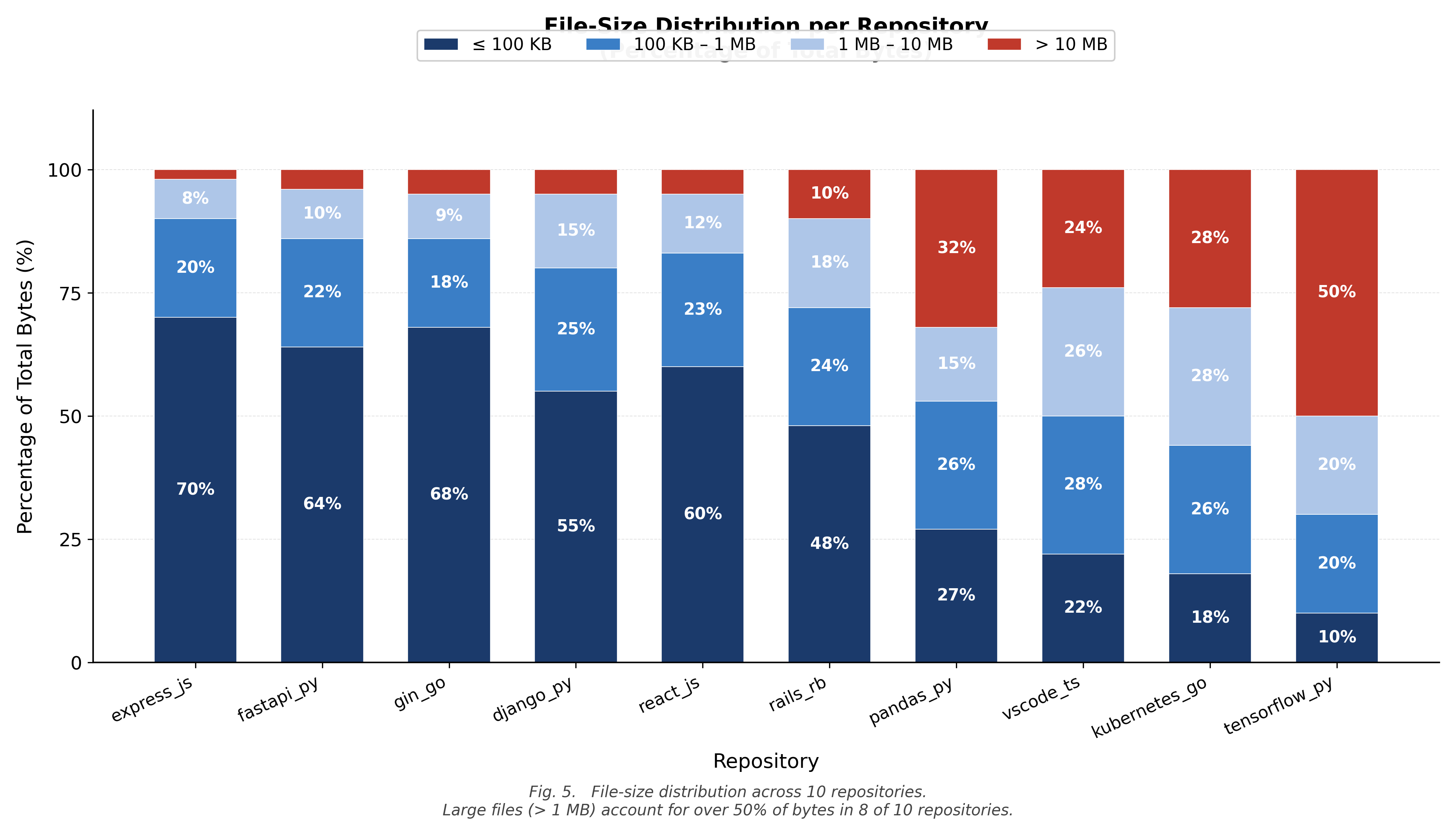}
  \caption{File-size distribution per repository as a percentage of
  total bytes across four size buckets ($\leq$100\,KB, 100\,KB--1\,MB,
  1\,MB--10\,MB, $>$10\,MB). The tail-at-scale structure is pronounced:
  in \texttt{tensorflow\_py} and \texttt{pandas\_py}, files $>$1\,MB
  dominate the byte distribution despite representing fewer than 2\%
  of file count. This structural concentration is why a simple size
  threshold achieves near-maximum token reduction in data-heavy
  repositories.}
  \label{fig:a3}
\end{figure*}

\begin{figure*}[p]
  \centering
  \includegraphics[width=0.88\textwidth,height=0.80\textheight,keepaspectratio]{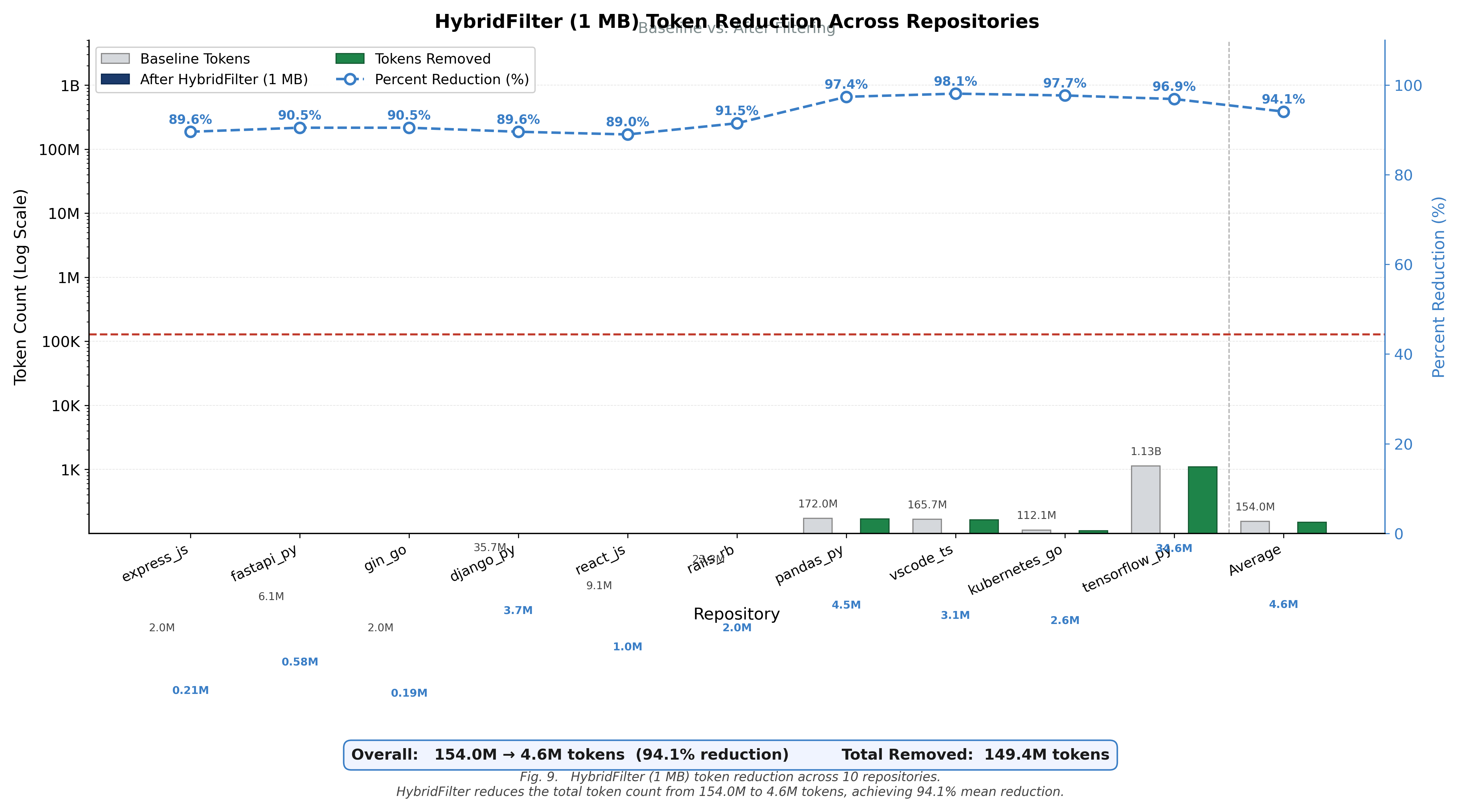}
  \caption{HybridFilter~(1\,MB) token reduction across all 10
  repositories plus corpus average. Grey bars: baseline token counts.
  Navy bars: token counts after filtering. Green bars: tokens removed.
  Blue line with right axis: percentage reduction per repository.
  Overall: 154.0\,M~$\to$~4.6\,M tokens (94.1\% aggregate reduction,
  149.4\,M tokens removed). This full-page version is provided for
  readability at reduced zoom and in print.}
  \label{fig:a4}
\end{figure*}


\begin{thebibliography}{00}

\bibitem{rag}
P.~Lewis \emph{et al.}, ``Retrieval-Augmented Generation for
Knowledge-Intensive NLP Tasks,'' \emph{NeurIPS}, vol.~33,
pp.~9459--9474, 2020.

\bibitem{brownfewshot}
T.~Brown \emph{et al.}, ``Language Models are Few-Shot Learners,''
\emph{NeurIPS}, vol.~33, pp.~1877--1901, 2020.

\bibitem{codex}
M.~Chen \emph{et al.}, ``Evaluating Large Language Models Trained on
Code,'' \emph{arXiv}:2107.03374, Jul.~2021.

\bibitem{tailscale}
J.~Dean and L.~A.~Barroso, ``The Tail at Scale,''
\emph{Commun.\ ACM}, vol.~56, no.~2, pp.~74--80, Feb.~2013.

\bibitem{llmlingua}
H.~Jiang \emph{et al.}, ``LLMLingua: Compressing Prompts for
Accelerated Inference,'' \emph{EMNLP}, pp.~13358--13376, 2023.

\bibitem{recomp}
F.~Pan, S.~Mallick, and T.~Rekatsinas, ``RECOMP: Improving
Retrieval-Augmented LMs with Context Compression,'' \emph{ICLR}, 2024.

\bibitem{sweagent}
J.~Yang \emph{et al.}, ``SWE-agent: Agent-Computer Interfaces Enable
Automated Software Engineering,'' \emph{NeurIPS}, 2024.

\bibitem{codellama}
B.~Roziere \emph{et al.}, ``Code Llama: Open Foundation Models for
Code,'' \emph{arXiv}:2308.12950, Aug.~2023.

\bibitem{swebench}
C.~E.~Jimenez \emph{et al.}, ``SWE-bench: Can Language Models Resolve
Real-World GitHub Issues?'' \emph{ICLR}, 2024.

\bibitem{copilot}
GitHub, ``GitHub Copilot---Your AI Pair Programmer,''
\url{https://github.com/features/copilot}, Apr.~2025.

\bibitem{tiktoken}
OpenAI, ``tiktoken: Fast BPE Tokeniser,''
\url{https://github.com/openai/tiktoken}, 2023.

\bibitem{paulsen2025}
N.~Paulsen, ``Context Is What You Need: The Maximum Effective Context
Window for Real World Limits of LLMs,''
\emph{arXiv}:2509.21361, Sep.~2025.

\bibitem{shi2023}
F.~Shi \emph{et al.}, ``Large Language Models Can Be Easily Distracted
by Irrelevant Context,'' \emph{ICML}, 2023.

\bibitem{copilotbilling}
GitHub, ``GitHub Copilot Transitions to AI Credits Usage-Based
Billing,'' May~2026.

\bibitem{acc}
C.~Smith and J.~Park, ``Active Context Compression: Autonomous Memory
Management in LLM Agents,'' \emph{arXiv}:2601.07190, Jan.~2026.

\bibitem{cmv}
R.~Thompson, ``Contextual Memory Virtualisation,''
\emph{arXiv}:2602.22402, Feb.~2026.

\bibitem{cursorrules}
S.~Jiang and D.~Nam, ``Beyond the Prompt: An Empirical Study of Cursor
Rules,'' \emph{MSR}, 2026.

\bibitem{housurvey}
X.~Hou \emph{et al.}, ``Large Language Models for Software Engineering:
A Systematic Literature Review,'' \emph{arXiv}:2308.10620, 2024.

\bibitem{gemfilter}
D.~Jin \emph{et al.}, ``GemFilter: Discovering Gems in Early Layers
for Accelerated Long-Context LLMs,''
\emph{arXiv}:2409.17422, Sep.~2024.

\bibitem{lostmiddle}
N.~F.~Liu \emph{et al.}, ``Lost in the Middle: How Language Models Use
Long Contexts,'' \emph{Trans.\ ACL}, vol.~12, 2024.

\bibitem{repocoder}
F.~Zhuo \emph{et al.}, ``RepoCoder: Repository-Level Code Completion
Through Iterative Retrieval and Generation,''
\emph{EMNLP}, pp.~2471--2484, 2023.

\bibitem{graphrag}
E.~S.~Edge \emph{et al.}, ``From Local to Global: A Graph RAG Approach
to Query-Focused Summarization,''
\emph{arXiv}:2404.16130, Apr.~2024.

\bibitem{treesitter}
M.~Brunsfeld \emph{et al.}, ``Tree-sitter: An Incremental Parsing
System for Programming Tools,''
\url{https://github.com/tree-sitter/tree-sitter}, 2018.

\bibitem{codebert}
Z.~Feng \emph{et al.}, ``CodeBERT: A Pre-Trained Model for Programming
and Natural Languages,'' \emph{EMNLP (Findings)}, pp.~1536--1547,
2020.

\end{thebibliography}
\end{document}